\documentstyle[multicol,aps,prl,epsf]{revtex}

\begin{document}

\draft

\title{Conductance Bistability of Gold Nano-wire at Room Temperature}

\author{Manabu Kiguchi$^1$, Tatsuya Konishi$^1$, and Kei Murakoshi$^{1,2}$}

\address{$^1$Department of Chemistry, Graduate School of Science, 
Hokkaido University, Sapporo, Hokkaido 060-0810, Japan}
\address{$^2$Precursory Research for Embryonic Science and Technology (PRESTO), 
Japan Science and Technology Agency, Sapporo, Hokkaido 060-0810, Japan JST PRESTO}

\date{\today}

\maketitle

\begin{abstract}

Quantized conductance behavior of gold nano wires was studied under electrochemical potential 
control. We fabricated 1 nm long mono atomic wires in solution at room temperature. 
Electrochemical potential significantly affected the stability of the mono atomic wire and fractional conductance 
peak occurrence in the conductance histogram. We revealed that the hydrogen adsorption on gold 
mono atomic wires was a decisive factor of the fractional peak, which was originated from 
the dynamic structural transition between two bistable states of the mono atomic wire showing the 
unit and the fractional values of the conductance. We could tune the stability of these bistable states 
to make the fractional conductance state preferable.

\end{abstract}

\medskip

\pacs{PACS numbers:  73.63.Rt, 73.40.Cg, 73.40.Jn}

\begin{multicols}{2}
\narrowtext

\section{INTRODUCTION}
\label{sec1}

Fabrication of stable atomic scale materials with quantized properties at room temperature 
is a central issue of material science. As one of the promising systems, gold nano wires have 
attracted wide attention. The conductance of the gold nano wire is often clearly quantized in the 
units of $G_{0}$ (1 $G_{0}$ = $2e^{2}/h$ =12.9 k$\Omega$$^{-1}$) \cite{pr377,prl81}. 
Successful preparations of a mono atomic gold wire 
have been reported with a scanning tunneling microscope (STM) or a mechanically controllable 
break junction (MCBJ)\cite{na395,na395-2}. While quantized conductance behaviour of the gold nano wire with 
the ideal unit of $G_{0}$  seems to be well-established in certain cases, the appearance of abnormal 
fractional conductance in the gold nano wire has recently attracted much attention 
\cite{prl91-2,prl80,prl90,prb60,nano4,prb67,prb55-2,zpb104,apl76,prb56,prl84}. Several 
models such as spin polarized conductance \cite{prl91-2,prl80}, structural deformation \cite{prl90,prb60,nano4,prb67}, 
scattering on impurities and defects \cite{prb55-2,zpb104,apl76}, and change in 
the electronic states \cite{prb56,prl84} have been proposed to explain the behaviour of fractional 
conductance. These models assume that the origins of fractional conductance are deviation of the 
one-dimensional (1D) mono atomic wire from the ideal structure and the possible contribution of 
Peierls transition \cite{prl90}. If one can assume the formation of the gold mono atomic wire, the structural 
transition between bistable states with distinct values of conductance, e.g. Peierls transition, can 
also contribute to well-defined fractional values at the conductance quantization. Although the 
origin of the fractional conductance has not been clarified yet, it seems to be very important to 
control the quantized behaviour of 1D materials by external perturbation, especially for the control 
of the reversible transition between the two states (1 $G_{0}$  and fractional conductance). 

While there are several external perturbations, such as temperature, pressure, electric and 
magnetic fields, etc, we paid attention to the electrochemical potential. In contrast to other external 
perturbations, the electrochemical potential can tune precisely the absolute potential energy of 
electrons in the nano wire. Recently, successful observations of conductance quantization at nano 
wires under electrochemical potential control have been reported for Au, Ag, Pb, Pd, Pt, Ni etc 
\cite{prl84,apl81,apl87,prb71,trans}. Fractional conductance was also observed by Xuo {\it et al.}  
for the gold nano wire in aqueous 
solution \cite{prl84}. However, the origin of the fractional conductance has not been clarified yet. 

To evaluate the origin of the fractional conductance observed in the gold nano wire, it is 
necessary to clarify the decisive factors. The measurement of dynamic changes in conductance is 
also important. In the present study, we studied conductance quantization of the gold nano wire at 
room temperature under electrochemical potential control.  The stability of the nano wire with 
fractional conductance was successfully controlled by the electrochemical potential. The origin of 
the fractional value was discussed based on the analysis of the time dependent change in the 
conductance values at the stretched process of the nano wires. 

\section{EXPERIMENTAL}
\label{sec2}

The experimental design used in this study was the same as described in detail in our previous 
report \cite{trans}. The experiments were performed in a four-electrode electrochemical cell mounted in a 
chamber that was filled with high purity N$_{2}$ gas to reduce the effect of oxygen. The gold nano wire 
was created by driving a STM tip in and out of contact with a gold substrate at a typical rate of 50 
nm/s in the electrochemical cell. The tip was made of a gold wire (diameter 0.25 mm) coated with 
wax to eliminate ionic conduction. The gold substrate was Au(111) prepared by the frame annealing 
and quenching method. The electrochemical potential ($\phi$) of the gold substrate and the tip was 
controlled using a potentiostat (Pico-Stat, Molecular Imaging Co.) with a Ag/AgCl reference 
electrode. A 0.50 mm diameter Pt wire was used as a counter electrode. The electrolyte was 0.1M 
Na$_{2}$SO$_{4}$ or 50 mM H$_{2}$SO$_{4}$. Conductance of the gold nano wire was calculated on the basis of the 
observed current between the tip and the substrate at the potential difference of 20 mV.

\section{RESULTS AND DISCUSSION}
\label{sec3}

\begin{figure}
\begin{center}
\leavevmode\epsfysize=60mm \epsfbox{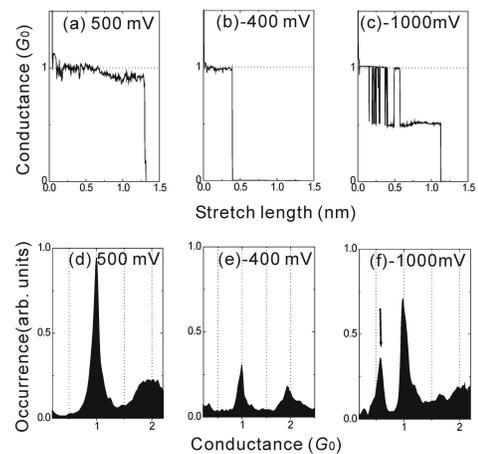}
\caption{
(a-c) Conductance trace and (d-f) conductance histogram of gold nano wires in 0.1 M 
Na$_{2}$SO$_{4}$  at electrochemical potential of (a,d) +500 mV, (b,e) -400 mV, and (c,f) -1000 mV. }
\label{fig1}
\end{center}
\end{figure}

\begin{figure}
\begin{center}
\leavevmode\epsfysize=50mm \epsfbox{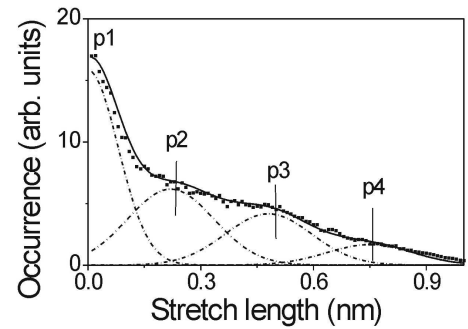}
\caption{
The distribution of lengths for the last conductance plateau for Au. The length of the last 
plateau was defined as the distance between the points at which the conductance dropped below 
1.3 $G_{0}$ and 0.7 $G_{0}$, respectively. The dot: experimental data, line: fitting results with distinct four 
Gaussian functions (dotted line).}
\label{fig2}
\end{center}
\end{figure}

Figure~\ref{fig1} shows the typical conductance traces and the conductance histogram of gold nano 
wires observed at $\phi$ = +500, -400, and -1000 mV in 0.1 M Na$_{2}$SO$_{4}$ . Each conductance histogram 
was obtained for a large number (over 3000) of individual conductance traces. The plateau of 1 $G_{0}$ 
stretched 1 nm in length at +500 mV (Fig.~\ref{fig1}(a)). Figure~\ref{fig2} shows the distribution of lengths for the 
last conductance plateau for Au. The length of the last plateau was defined as the distance between 
the points at which the conductance dropped below 1.3 $G_{0}$ and 0.7 $G_{0}$, respectively. 
The experimental results (dot in the figure) could be fitted with four 
distinct Gaussian functions. The values of the Gaussian peaks located at 0.00, 0.22, 0.48 and 0.75 
nm, respectively. The inter-peak distances were about 0.25 nm, that were in agreement with the 
calculated Au-Au distance in the gold mono atomic wire \cite{pr377}. Since it has been shown that the 
conductance of the mono atomic gold contact is 1 $G_{0}$  \cite{pr377,prl81}, 
the 1 nm long 1 $G_{0}$  plateau in Fig.~\ref{fig1} and 
the length histogram in Fig.~\ref{fig2} proved the formation of the gold mono atomic wire in the present 
system. While a 2 nm long gold mono atomic wire is fabricated at 4.2 K in ultra high vacuum 
(UHV) \cite{na395-2}, there are few reports of the formation of a mono atomic wire at room temperature. The 
present results proved that the electrochemical method made it possible to fabricate a stable gold 
mono atomic wire at room temperature in solution. 

\begin{figure}
\begin{center}
\leavevmode\epsfysize=50mm \epsfbox{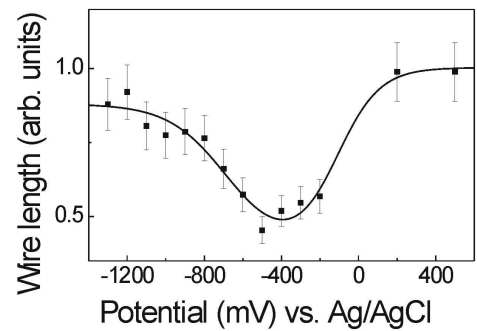}
\caption{Electrochemical potential dependence of the averaged stretch length of the gold mono 
atomic wire.}
\label{fig3}
\end{center}
\end{figure}

The length of the conductance plateau was dependent upon the electrochemical potential. 
Figure~\ref{fig3} shows the averaged stretch length of the gold mono atomic wire as a function of $\phi$. As the 
potential was scanned from 500 mV to negative, the length decreased at 200 mV, and reached a 
minimum value at -400 mV. Polarization more negative than -600 mV led to the recovery of the 
length. At a potential regime more positive than $\phi$= -400 mV, sulfate anions adsorb on the surface 
of gold. Gradual increase in the length from -400 mV to 500 mV agreed with previously 
reported increments in the amount of the adsorbed sulfate anions on the gold electrode \cite{jec452}. This 
result suggested that adsorption of the anions onto the gold nano wire led to stabilization of the gold 
mono atomic wire. Polarization to a potential more negative than -400 mV induces hydrogen 
evolution reaction at the gold electrode \cite{ea31}. Thus, recovery of the stretch length at the negative 
potential regime implied that evolved hydrogen molecules and/or adsorbed hydrogen could stabilize 
the gold mono atomic wire. These electrochemical potential dependent changes in the stretch length 
were fully reversible. This fact implies that the stability of the gold mono atomic wire could be 
controlled by the electrochemical potential. 

\begin{figure}
\begin{center}
\leavevmode\epsfysize=30mm \epsfbox{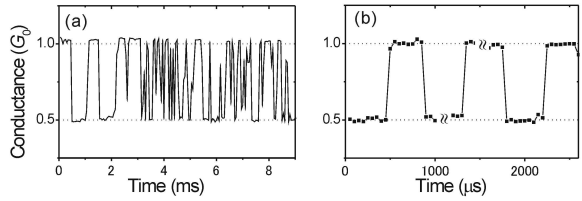}
\caption{Time dependent change in the conductance of the gold nano wire at $\phi$ = -1000 mV. Data 
were acquired every 50 $\mu$sec.}
\label{fig4}
\end{center}
\end{figure}

At the negative potential regime, reversible transition of the conductance between 1 $G_{0}$ and 
0.5-0.7 $G_{0}$ was observed as shown in Fig.~\ref{fig1}(c). Extended conductance traces shown in Fig.~\ref{fig4} proved 
that the changes between the two states occurred within 50 $\mu$sec. This conductance fluctuation 
resulted in the fractional conductance peaks in the conductance histogram (see Fig.~\ref{fig1} (f)). The 
conductance fluctuated between two well-defined values, which strongly suggested the occurrence 
of a structural transition between two bistable states at the gold mono atomic wire. In solution, 
hydrogen evolution proceeds at the electrochemical potential in which the fractional value was 
observed. Therefore, hydrogen molecules and/or adsorbed hydrogen atoms may play a decisive role 
in the evolution of the fractional conductance. 

\begin{figure}
\begin{center}
\leavevmode\epsfysize=60mm \epsfbox{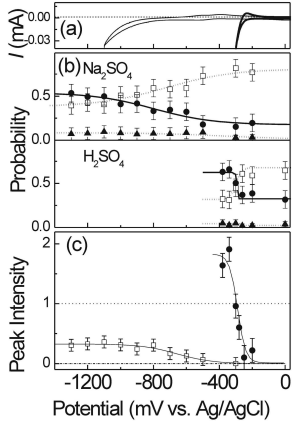}
\caption{(a) Voltammogram in 0.1 M Na$_{2}$SO$_{4}$  (thin line) and 50mM H$_{2}$SO$_{4}$  (bold line). (b) The 
probability of each conductance trace in the presence of only fractional conductance step (triangle), 
only 1 $G_{0}$ conductance step (box), and both steps (filled-circle) as a function of electrochemical 
potential($\phi$). (c) The normalized fractional conductance intensity as a function of $\phi$ in Na$_{2}$SO$_{4}$  (box) 
and H$_{2}$SO$_{4}$  (filled-circle). }
\label{fig5}
\end{center}
\end{figure}

Figure~\ref{fig5} shows potential dependences of the reductive current due to the hydrogen evolution 
together with those of the occurrence of the fractional values in 0.1 M Na$_{2}$SO$_{4}$ . Figure ~\ref{fig5}(b) shows 
the probability of the conductance trace in the presence of only a fractional conductance step, only a 
1 $G_{0}$ conductance step, and both steps as a function of $\phi$. Figure~\ref{fig5}(c) shows the intensity of the 
fractional conductance peak normalized by the 1 $G_{0}$ peak. With an increase in current due to the 
hydrogen evolution reaction, the probability of the conductance trace with the fractional 
conductance steps and the intensity of the fractional conductance peak increased. In the presence of 
the fractional conductance, both 1 $G_{0}$ and fractional conductance steps simultaneously appeared, 
suggesting that the fractional value originated from the conductance fluctuation. These results 
showed that evolved hydrogen induced the fractional conductance, which may reflect the 
occurrence of the transition between two bistable states of the gold mono atomic wire at the 
negative potential regime.

\begin{figure}
\begin{center}
\leavevmode\epsfysize=40mm \epsfbox{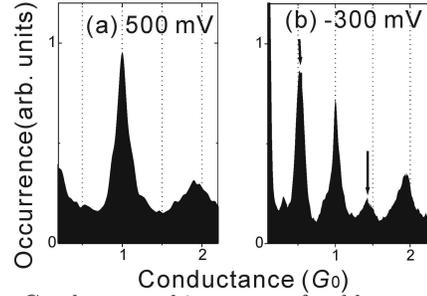}
\caption{Conductance histogram of gold nano wires in 50 mM H$_{2}$SO$_{4}$  at electrochemical potential 
of (a)+500 mV and (b) -300 mV.}
\label{fig6}
\end{center}
\end{figure}

To confirm the effect of hydrogen on the fractional value, we studied electrochemical potential 
dependence of the conductance quantization behavior in acidic solution. Figure~\ref{fig6} shows the 
conductance histogram of gold nano wires observed at $\phi$ = +500 and -300 mV in 50mM H$_{2}$SO$_{4}$. The 
fractional conductance peak appeared at a potential regime more negative than $\phi$ = -200 mV. It is 
noteworthy that the intensity of the fractional conductance peak was larger than that of 1 $G_{0}$ peak. 
We discuss this point later. In the solution with a lower pH (50 mM H$_{2}$SO$_{4}$), the onset potential of 
the hydrogen evolution shifted to more positive potential than that which occurred in the solution 
with a neutral pH (0.1 M Na$_{2}$SO$_{4}$), as shown in Fig.~\ref{fig5}(a). The occurrence of the fractional 
conductance shifted in accordance with that of the hydrogen evolution (see Fig.~\ref{fig5}(c)). Dependence 
of the solution pH confirmed that the fractional conductance observed in solution originated from 
the structure stabilized by hydrogen. 

Recently, similar quantized conductance behavior was reported by Csonka {\it et al.} \cite{prl90} They 
observed a fractional conductance peak (0.6 $G_{0}$) in a conductance histogram of gold nano wires 
under hydrogen dosing at 20 K in UHV, although the intensity of the fractional feature was very 
weak compared with that of the 1 $G_{0}$ peak. They observed reversible conductance fluctuation 
between 1 $G_{0}$ and fractional conductance in the conductance trace. They proposed that the fractional 
conductance value observed at the conductance fluctuation originated from the structural transition 
between an equal-spacing and ta dimerized wire, based on the following reasons. First, the 
fractional conductance peaks were not observed in Cu and Ag, and Au forms a mono atomic wire, 
while Cu and Ag do not form a mono atomic wire \cite{pr377}. Second, a stretched single atom wire has a 
strong tendency to spontaneous dimerization - a predecessor for Peierls transition in extended 1D 
systems \cite{pi}. Theoretical calculation supports that the conductance of a dimerized gold wire is 
0.4-0.6 $G_{0}$ \cite{prb60}, which is close to the observed values. Since conductance quantization behavior in 
UHV was similar to that observed in the present study, a similar phenomena would occur in both 
UHV and solution. The fractional conductance and conductance fluctuation observed in the present 
study might originate from the dimerized wire and the structural transition between equal-spacing 
and the dimerized wire. If the dimerization is the origin of the fractional conductance, we could 
observe the Peierls transition of the 1D metal (mono atomic wire) at room temperature. 

At the present stage, there is no direct proof of the reversible structural transition of the gold 
mono atomic wire in solution. As proposed by Csonka {\it et al.}, the apparent effect of the hydrogen 
evolution reaction on the appearance of the fractional conductance value strongly suggests that 
adsorbed hydrogen molecules on the gold mono atomic wire could contribute to the formation of 
two bistable states of the mono atomic wire.
Another possibility for the formation of the hydrogen incorporated wire can be 
considered, because a recent theoretical calculation showed that the incorporation of a hydrogen 
molecule in the gold mono atomic wire also decreased conductance below 1 $G_{0}$\cite{nano4,prb67}. Further 
investigation is needed to determine the structure with fractional conductance stabilized by 
hydrogen.

The present results clearly demonstrate the fabrication of a well-defined atomic structure with 
fractional conductance values controlled by the electrochemical potential \cite{cond}. In contrast with the 
UHV case, we controlled the intensity of the fractional conductance peak (stability) by the 
electrochemical potential. Furthermore, the relative intensity of the fractional peak became stronger 
than that of the unit value (see Fig.~\ref{fig6}). The specific structure with 0.5-0.7 $G_{0}$ (fractional wire) 
became more stable than the structure with 1 $G_{0}$ (normal wire). It should be noted that the intensity 
of the fractional conductance feature does not exceed the 1 $G_{0}$ peak intensity even in the system in 
UHV at 20 K \cite{prl90}, that is, fractional wire cannot be kept as the most energetically favorable 
structure. The present observation proved that structural bistability of gold mono atomic wire could 
be tuned by changing the electrochemical potential.

Recently, several atomic quantum switches have attracted wide attention as a conceptual new 
electric switch \cite{prl93}. In the atomic quantum switch, a contacting metal atom is electrochemically 
deposited (on) or dissolved (off) in the contact area. On the other hand, in the present case, the 
conductance transition was based on the transition between two bistable structures, and thus, the 
movement of metal atom would be very small. Furthermore, both stability and structure of the 
mono atomic wire can be controlled at room temperature. Therefore, the present findings would 
also open an intriguing perspective for the emerging fields of quantum electronics and logics on the 
atomic scale in the future. 

\section{CONCLUSIONS}
\label{sec4}

We have studied quantum conductance behaviour of the gold nano wire under electrochemical 
potential control. A stable 1 nm long mono atomic wire was formed in solution, and a certain atomic 
configuration of the wire with fractional conductance was stabilized by hydrogen. The 
conductance bistability of the gold mono atomic wire could be controlled by the electrochemical 
potential and the pH at room temperature. 

\acknowledgments{
This work was partially supported by a Grant-in-Aid for Scientific Research A (No. 16205026) 
and Grant-in-Aid for Scientific Research on Priority Areas (No. 17069001) from MEXT.}

\end{multicols}
\end{document}